\documentclass{llncs}[]
\usepackage{amssymb}
\usepackage{graphicx}
\usepackage{url}
\pdfoutput=1

\begin{document}

\title{Towards platform-independent verification \\ of the standard mathematical functions: \\
the square root function\thanks{This research is supported by Russian Basic Research Foundation grant
no. 17-01-00789 \emph{Platform-independent approach to formal specification and verification of standard mathematical functions}.}}

\titlerunning{Towards verification of the square root}

\author{Nikolay V. Shilov\inst{1}, Igor S. Anureev\inst{2}, Mikhail Berdyshev\inst{1}, Dmitry Kondratev\inst{2}, Aleksey V. Promsky\inst{2}}

\authorrunning{N. Shilov et. al.}

\institute{Innopolis University,
Innopolis, Russia\\ \email{shiloviis@mail.ru}, \email{m.berdyshev@innopolis.ru}
\and A.P. Ershov Institute of Informatics Systems RAS, Novosibirsk, Russia\\
\email{anureev@iis.nsk.su}, \email{apple-66@mail.ru}, \email{promsky@iis.nsk.su}}


\maketitle

\begin{abstract}
The paper presents (human-oriented) specification and (pen-and-paper) verification of the square root function.
The function implements Newton method and uses a look-up table for initial approximations.
Specification is done in terms of total correctness assertions with use of precise arithmetic and the mathematical square root $\sqrt{\dots}$,
algorithms are presented in pseudo-code with explicit distinction between precise and machine arithmetic,
verification is done in Floyd-Hoare style and adjustment (matching) of runs of algorithms with precise arithmetics and with machine arithmetics.
The primary purpose of the paper is to make explicit properties of the machine arithmetic that are sufficient
to make verification presented in the paper.
Computer-aided implementation and validation of the proofs (using some proof-assistant) is the topic for further studies.

\end{abstract}

\section{Introduction}\label{Intro}
\subsection{Motivation}
Let us start with a quotation from the abstract
of the paper \cite{Monniaux08}, because it correlates with the purpose of our
paper very well:
\begin{quote}
\emph{Current critical systems commonly use a lot of floating-point computations,
and thus the testing or static analysis of programs containing floating-point operators
has become a priority. However, correctly defining the semantics of common
implementations of floating-point is tricky, because semantics may change with many
factors beyond source-code level, such as choices made by compilers.}
\end{quote}
The major difference between \cite{Monniaux08} and the present paper is the concern:
the cited paper addresses problems
with the floating-point value representation and arithmetics,
while the present paper addresses the standard mathematical function
platform-independent formal specification and formal verification
by study in full details the square root function.

In our approach the platform-independence means
that we specify properties of the functions and prove these properties for
for approximate algorithms without building (some-how comprehensive) formal model of a particular processor architecture
(like, for example, Intel's processors in \cite{Harrison00a,Harrison00b,Harrison03} or Oracle’s processors in \cite{Grohoski17})
or fix/floating-point formats (like, for example, in \cite{Barret89,Harrison99,Monniaux08})
but carry out a proofs with several explicit simple assumptions how machine arithmetic relates to the precise arithmetic.
Thus these explicit simple assumptions are sufficient conditions to validate on
a particular processor with particular formats of numeric data
in order to guarantee that a mathematical functions verified with these assumptions (square root function in this paper)
meet their formal specification.
We believe that our assumptions about machine arithmetic are valid for many platforms and
are easy to check/validate.

Before we move to a literature survey on topics related to our study
let us  advocate importance of the formal verification of the software.
Although our paper addresses verification \emph{in small} (i.e. verification of small stay-alone programs),
we would appeal in the next paragraph to importance of verification  \emph{in large} i.e. verification of complex cyber-physical systems.
(Please refer slides 8-11 of \cite{Grohoski17} for justification of a need of verification in small,
floating-point arithmetic, and the standard mathematical functions in particular.)

December 12, 2017, Roskosmos \cite{Roscosmos} has published the official
results of investigation of the accident on November 28, 2017, which has led to loss of
the \emph{Meteor-M} satellite (altogether with another 18 satellites).
Risks at start have been insured  for the sum of 2,6 billion Russian rubles.
Results of the investigation read \cite{Roscosmos} (translation by N.V. Shilov):
\begin{quote}
\emph{It has revealed the hidden problem in an algorithm which
wasn't shown for decades of successful launches of \emph{Sojuz} carrier-rocket with the upper-stage  accelerating block \emph{Fregate}.
...
There was a combination of parameters of a launching-pad
of the spaceport, azimuths of flight of the carrier-rocket
and the accelerating block which hadn't been met earlier.
Respectively, it hasn't been revealed at the carried-out
on-land testing and simulation of a ballistic trajectory according to the
standard adopted techniques.}
\end{quote}
The formal verification (in large as well as in small) is aimed to reveal ``hidden problem in an algorithm which
wasn't shown for decades'', a rare ``combination of parameters''
that can't be revealed ``according to the standard adopted techniques''.
We believe that a formal verification is a demand of the day and one of a few \emph{grad challenges} for Computer Science research \cite{Hoare03}.

\subsection{Literature survey}
A need for better specification and validation
of the standard functions is recognized (in principle)
by industrial and academic professional community,
as well as the problem of a conformance of their implementation with the
specification.
We would like to point out just on two papers \cite{Kuliamin07,Kuliamin10}
that address formal \emph{complex} specification and testing of standard mathematical functions.
Hear we use adjunctive \emph{complex} because these papers don't restrict function properties
by \emph{accuracy} but take into consideration, for example, that $\sin$ and $\cos$ are odd and even functions respectively,
they match pythagorean normalization equality $\sin^2x+\cos^2x=1$ for all real $x\in \mathbb{R}$.
An educational value and issues of better documentation and specification of the standard functions are discussed
in papers \cite{Shilov15,ShilovPromsky16}.

Several studies have been published on platform-dependant formal verification of mathematical functions,
including division \cite{FergusonBinghamErkokHarrisonLeslie-Hurd17,Harrison00b},
square root \cite{SawadaGamboa02,Harrison00b,Shelehov10,FergusonBinghamErkokHarrisonLeslie-Hurd17},
trigonometric \cite{Harrison00a},  exponential \cite{Harrison00c}, and gamma \cite{SiddiqueHasan14} functions.
Also several studies have been published on axiomatization of machine arithmetic
(mostly binary floating-point arithmetic for the IEEE-754 standard \cite{IEEEstandard})
to prove basic mathematical properties and consequently prove correctness of mathematical functions
\cite{Barret89,Harrison99,BrainTinelliRuemmerWahl15,AyadMarche10}.

First let us remark that even platform-independent verification of the integer square root function is not a trivial exercise.
Please refer, for example, paper \cite{Shelehov10} where some standard mathematical integer functions (including the square root)
are specified and verified in PVS.

Paper \cite{Barret89} formalizes machine arithmetic using Z-notation and present an implementation of the specification written in Occam.
It presents a formal description of several mathematical functions over floating-point numbers, namely:
rounding, addition, multiplication, square root, type-casting to integer, comparisons, etc.
Besides it, the paper specifies five classes of floating-point numbers: NaN, Inf, zero, normal and denormal numbers.
Then four modes of rounding and error conditions are presented.
The implementation includes representations of floating-point numbers,
its rounding and packing/unpacking and basic finite mathematical procedures.
The main algorithm pattern for a binary operation with floating-point values (according to \cite{Barret89}) is as follows:
\begin{enumerate}
\item unpack both operands into their sign, exponent and mantissa fields;
\item denormalise both by shifting in the leading bit of the mantissa if necessary;
\item perform the  operation with denormalized arguments;
\item pack the result and then round the packed result.
\end{enumerate}

Next paper \cite{Harrison99} presents an approach to verification in HOL Light of several floating-point operations of a new (at time of publication)
Intel computer architecture IA-64:
\begin{quote}
\emph{Correctness of the mathematical software starts from the assumption that the
underlying hardware floating point operations behave according to the IEEE standard 754
for binary floating point arithmetic.
Actually, IEEE-754 doesn't explicitly address floating-point machine arithmetic operations,
and it leaves underspecified certain significant questions,
e.g. NaN propagation and underflow detection.
Thus, we not only need to specify the key IEEE concepts but also some details specific to IA-64.}
\end{quote}
This paper starts with a theory of floating point arithmetic, which is non specific to any format
and afterwards specifies IA-64 formats in details.
Floating-point numbers in \cite{Harrison99} are presented in highly generic way (as $\pm k\times 2^{E-N}$)
but have a canonical representation and normalized form.
The paper argues also that the concept of the unit in the last place (ulp) has several different definitions but
all have some counterintuitive properties; due to this reason the paper adopts a modified definition from \cite{Muller05}.
Four types of rounding (to-Nearest, Down, Up, to-Zero) are defined in \cite{Harrison99};
in contrast to the IEEE standard rounding is defined for numbers with an unbounded exponent range, but
all overflows are handled during operations execution.

Paper \cite{BrainTinelliRuemmerWahl15} describes syntax and semantics of floating-point arithmetic theory.
Besides being general, the formalization seriously rely upon Satisfiability Modulo Theories (SMT) approach.
The paper has 2 certain contributions: mathematical structures for floating-point model,
a signature for a theory of floating-point arithmetic and an interpretation of its operators
in terms of the mathematical structures defined earlier.
Thus, it is designed to be a formal reference for automatic theorem provers providing built-in support for reasoning about floating-point arithmetic.

It is important to prove correctness of mathematical functions widely used in different architectures and libraries.
The square root function is required (by IEEE-754) to be \emph{exact}
(please refer  the next section \ref{WhatIsSQRT} for the definition).
Hence, correctness of this function (as well as other exact functions) should be considered with a special attention.

Approach to verification of the square root function suggested in paper \cite{FergusonBinghamErkokHarrisonLeslie-Hurd17}
is based on a concept of a digit serial method (DSM) for a number:
DSM for a real number $x\in \mathbb{R}$ is an algorithm that determines the digits of $x$ serially, starting with the leading digit.
The main contribution of the paper is a generic DSM analysis method for determining bounds on the magnitudes of the digits,
as well as bounds on the error associated with the estimates.
(We believe that the approach may be related to interval techniques \cite{Gutowski}.)

In the papers \cite{Gamboa97,SawadaGamboa02} authors prove correctness of the square root algorithm used in Power4 processor.
The algorithm uses Chebyshev polynomials.
Despite of the fact that the algorithm has more steps comparing to the Newton (also known as Newton-Raphson) method used in \cite{Harrison03},
only one iteration is enough to get necessary accuracy;
also, because less instructions are reliant on the earlier ones in the polynomial algorithm, the algorithm is better for parallelization.
The verification in \cite{Gamboa97,SawadaGamboa02}
is divided by two parts: proof of Taylor’s theorem and proof of properties of the square root function using Taylor’s theorem.
One of the biggest challenges in the study in \cite{SawadaGamboa02} was to approximate error size of Chebyshev polynomial
with Taylor’s series as the former has a better approximation. To escape this problem hundreds of Taylor’s series were evaluated.
The proof has been carried-out using non-standard analysis book (library) in ACL2.

\subsection{Paper structure}
In the next section \ref{WhatIsSQRT} we present specification of the square root function according to the C programming language standard,
sketch Newton method to compute approximations for the square root, formalize it as $SQR$ algorithm (with until-loop)
and specify it in Floyd-Hoare style by a total correctness assertion \cite{Gries81}
assuming the precise arithmetic (i.e. for mathematical reals).

In section \ref{PenPapSQRT} we give a pen-and-paper verification of the algorithm $SQR$ from the previous section \ref{WhatIsSQRT}
using  Floyd-Hoare approach \cite{Gries81} and assuming the precise arithmetic:
partial correctness is considered in the subsection \ref{PartCorSQR} and termination --- in the subsection \ref{TermSQR}.

Section \ref{TowMachSqrt} presents two modifications of the square root algorithm $SQR$:
the first algorithm $ISQR$ differs from $SQR$ by use of an auxiliary function to ``compute''
good initial approximations (see subsection \ref{impSQRT}),
the second algorithm $FSQR$ (see subsection \ref{forSQRT}) is a for-loop-based algorithm that uses the same auxiliary function but
(in contrast to $ISQR$) estimates the number of sufficient iterations to achieve the required accuracy of the approximations.
Both algorithms in this section are specified and verified under assumption that the arithmetic is precise.

The following-up section \ref{fixSQR} starts with the subsection \ref{FixPointArith}
where we formulate assumptions about fix-point values and arithmetic, and then presents and specifies
the fix-point algorithm $fixSQR$ in the subsection \ref{SpecfixSQRT}.

The algorithm $fixSQR$ is verified (manually) in the section \ref{verFixSQR} by comparison
with runs of algorithm $FSQR$ on the same input data. In the same section \ref{fixSQR} we specialise the algorithm $fixSQR$
into better algorithm $mixSQR$ which is correct because of correctness of the algorithm $fixSQR$.

Section \ref{fltSect} presents our assumptions about floating-point arithmetic, the algorithm $fltSQR$ that computes
approximations for the square root function in floating-point arithmetic,
its specification and pen-and-paper verification.
The algorithm is based on square root extraction from mantissa (using the fix-point algorithm $mixSQR$)
and integer division to compute the exponent.

In the last section \ref{Concl} we summarise the content and contribution of the present paper and
discuss the topics for further research.

\section{What is the standard function \texttt{sqrt}? }\label{WhatIsSQRT}
The C reference portal  at \url{en.cppreference.com/w/c}
specifics the the square root function \texttt{sqrt} \cite{SqrtCref} as it represented in the Appendix \ref{sqrt-spec}.
It is easy to see an ambiguity in the specification:
it first says that \texttt{sqrt(2)} must be $\sqrt{2}$, but then (in the Notes) that
the error of \texttt{sqrt(2)} must be less than $0.5$ of \emph{ulp} --- the unit in the
least precision (that is type and platform dependable.
Of course, we have to rule out the first option (that \texttt{sqrt()} is $\sqrt{\ }$)
as non-realistic;
instead we have and examine in details the second one.

The standards mentioned in the specification are
IEEE 754-2008 Standard for Floating-Point Arithmetic
and the international standard ISO/IEC 60559:2011 \cite{IECstandard}
(that is identical to IEEE 754-2008).
Section 9 of the standard recommends fifty operations
that language standards should define (but all these operations are optional,
not required in order to conform the standard).
Some of these operations (including \texttt{sqrt()} as a special case of the function
$(\ )^{1/n}$ for $n=2$),
if being implemented, must be (according to the standard' terminology) \emph{exact}
i.e. to round correctly (i.e. with an error less than $0.5$ulp).
Due to this use of the term \emph{exact} for computer functions and operation,
let us fix another term \emph{precise} when we speak about mathematical functions and operations
with mathematical real numbers $\mathbb{R}$.

The first problem with the standard is type and platform dependence of the concept of the exact function:
the accuracy upper bound $0.5$ulp depends on numeric type (float vs. double) as well as on implementation of the types
(i.e. memory size reserved for the types).
Another very critical problem with the specification and ISO/IEC/IEEE standards above
is the absence (in the specification and standards)
of a description of any validation procedure to check/prove that
an implementation conforms the specification/standard.

Instead of requiring that \texttt{sqrt} computes the exact values for square roots
in type- and/or platform-dependent way,
it makes sense to specify another ``standard''
\emph{generic} function (say $SQR(\ ,\ )$) for \emph{generic}
numeric data types with two parameters:
the first parameter is for passing the argument value $Y\geq 0$
and the second --- for passing the accuracy value $Eps>0$;
the function is for computing $\sqrt{Y}$ with the accuracy $Eps$.

The accuracy  of this function $SQR$
(i.e. the most wanted property and the only property specified in the standard)
can be formally specified by
any (or both) of the following two assertions:
\begin{itemize}	
\item for all type-legal values $y\geq 0$ and $\varepsilon>0$, $SQR(y, \varepsilon)$ differs from $\sqrt{y}$
by no more  than $\varepsilon$, i.e. $|\sqrt{y} - SQR(y,\varepsilon)|\leq \varepsilon$;
\item for all type-legal values $y\geq 0$ and $\varepsilon>0$, $\big(SQR(y, \varepsilon)\big)^2$ differs from $y$
by no more  than $\varepsilon$, i.e. $\big|y - (SQR(y, \varepsilon)\big)^2\big|\leq \varepsilon$.
\end{itemize}

It makes sense to fix the first formal specification
for better compatibility with the concept of the exact standard function,
since in this case we can define the standard function \texttt{sqrt}
via \texttt{SQR} as follows:
\begin{verbatim}
    float sqrt(float Y)
    {return((float)SQR(Y, default(float)/2.0);}
\end{verbatim}
where \texttt{default} is another new type- and platform-dependent feature
(similar to \texttt{sizeof})
that returns the value of the unit in the least precision for a numeric type.

One may select any reasonable and feasible computation method
to approximate $\sqrt{\ }$.
For example, it can be a very intuitive, easy-to-implement and popular in education (e.g. \cite{Muller05,Kochan05b}) Newton Method:
\begin{enumerate}
\item input the number (to compute the square root) and guess an initial approximation for the root;
\item \label{step2} compute the arithmetic mean between the guess and the number divided by the guess;
let this mean be a new guess;
\item repeat step \ref{step2} while
the difference between the new and the previous guesses isn't small enough (i.e. doesn't feet the use-defined accuracy).
\end{enumerate}
(Please refer to Fig. \ref{SQRimpl} for a sample implementation of the function
for the data type \texttt{float}.)
\begin{figure}[t]
\begin{verbatim}
float ab(float X)
{if (X<0) return(-X); else return(X);}

float SQR(float Y, float Eps)
{float X, D;
       X=Y;
       do {D=(Y/X-X)/2; X+=D;} while (ab(D)>=Eps/2);
return X;}
\end{verbatim}
    \caption{A floating-point function to compute
    a square root approximation}\label{SQRimpl}
\end{figure}

Both floating-point functions in Fig. \ref{SQRimpl} are easy to specify formally in a Hoare style \cite{Gries81}:
\begin{equation}\label{firstSpec}
\left\{\begin{array}{l}
  [X\ is\ float]\ ab(X)\ [returned\ value\ =|X|], \\

  [Y \geq 0\ and\ Eps>0\ are\ floats]  \\
 \hspace*{1cm} SQR(Y,\ Eps)[|returned\ value\ -\sqrt{Y}|\leq Eps].
\end{array}\right.
\end{equation}
(Remark  that
the specification is incomplete since it doesn't specify the program behavior and output
if input values are $Y<0$ and/or $Eps\leq 0$.)

For a generic square root function with a generic numeric data type for input and output values, the specification (\ref{firstSpec})
should be modified:
\begin{equation}\label{SQRspec}
\begin{array}{c}
  [ TYPY\ is\ a\ numeric\ type,\
  Y\geq 0:\ TYPE,\ and\ Eps>0:\ TYPE] \\
 \hspace*{\fill}  SQR(Y,\ Eps)\
 [|returned\ value\ -\sqrt{Y}|\leq Eps].
\end{array}
\end{equation}
If these specifications are proved, then \texttt{SQR} may be a good alternative
to the standard function \texttt{sqrt}.

Unfortunately, it is not easy to prove these specifications
automatically and formally because of several reasons.
The major one is a problem that we already discussed in the literature survey in the introduction section \ref{Intro} ---
an axiomatization of the computer-dependent floating-point arithmetic.
Even a manual pen-and-paper verification of the algorithm $SQR$
(assuming precise arithmetic for real numbers $\mathbb{R}$) is not a trivial exercise
that we solve in the next section.

\section{Pen-and-paper verification of $SQR$}\label{PenPapSQRT}
\begin{figure}[t]
\centering
  \includegraphics[width=8cm]{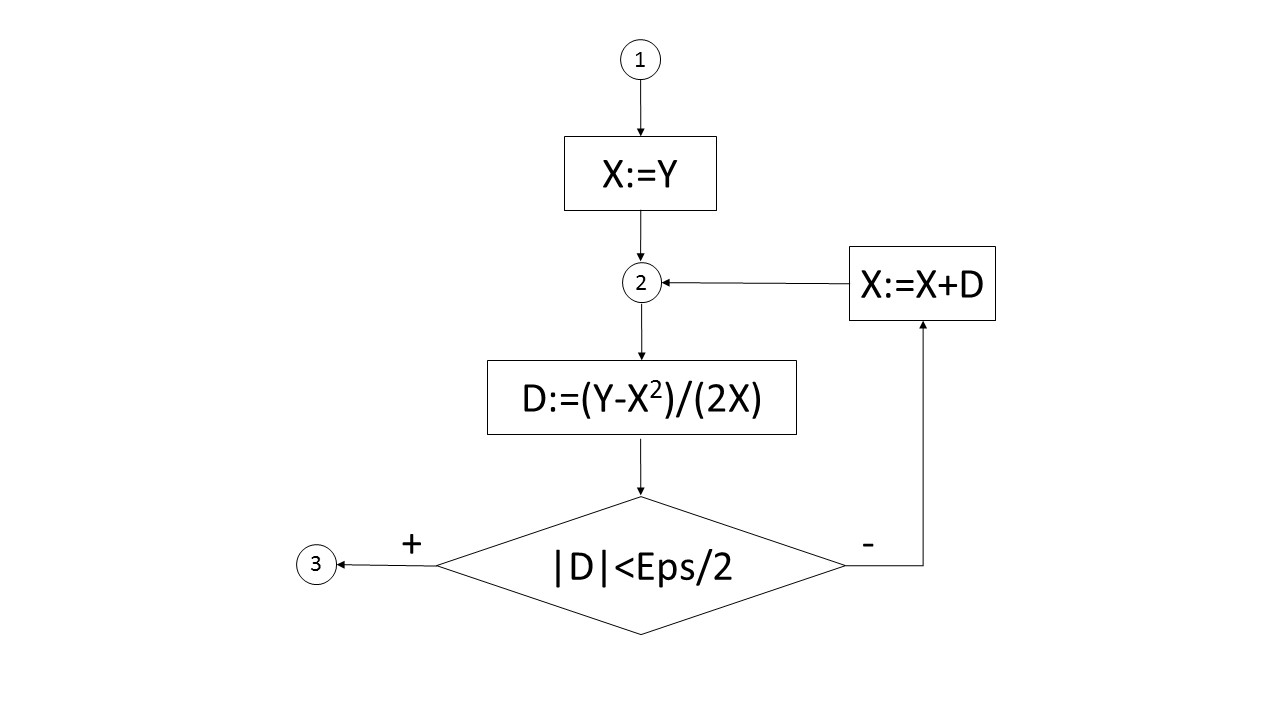}\\
  \caption{A flowchart of the algorithm $SQR$}
  \label{flowchart1}
\end{figure}
Fig. \ref{flowchart1}  shows a flowchart of
the algorithm (a little bit modified) of the function from Fig. \ref{SQRimpl}.
Let us refer to the algorithm as $SQR$ in the sequel.
Having specified the algorithm in the same way as the function, we need to prove
the following ``relaxation'' of the second triple in (\ref{SQRspec}):
\begin{equation}\label{SQRspecAll}
[Y,Eps\in\mathbb{R}\ \&\ Y\geq 0\ \&\ Eps>0]\ SQR\ [|X-\sqrt{Y}|\leq Eps]
\end{equation}
To prove this assertion, let us consider three disjoint cases for
the range of the initial value of the variable $Y$:
$0\leq Y<1$, $Y=1$ and $Y>1$:
\begin{equation}\label{SQRspec0}
[Y,E\in\mathbb{R}\ \&\ 0\leq Y<1\ \&\ Eps>0]\ SQR\ [|X-\sqrt{Y}|\leq Eps],
\end{equation}
\begin{equation}\label{SQRspec1}
[Y,E\in\mathbb{R}\ \&\ Y=1\ \&\ Eps>0]\ SQR\ [|X-\sqrt{Y}|\leq Eps],
\end{equation}
\begin{equation}\label{SQRspec2}
[Y,E\in\mathbb{R}\ \&\ Y>1\ \&\ Eps>0]\ SQR\ [|X-\sqrt{Y}|\leq Eps].
\end{equation}

The second case (\ref{SQRspec1}) is trivial.
Two other cases (\ref{SQRspec0}) and (\ref{SQRspec2}) are ``ideologically'' very similar,
so we prove below in this section the assertion (\ref{SQRspec2}) only.
Due to this reason we assume below in the subsections \ref{PartCorSQR} and \ref{TermSQR} that the initial (input) variable values
meet the precondition $Y,E\in\mathbb{R}\ \&\ Y>1\ \&\ Eps>0$
and that all operation used in the algorithm are precise mathematical operations with reals.

\subsection{Partial Correctness}\label{PartCorSQR}
Let us employ the Floyd method \cite{Gries81}
for a pen-and-paper proof of partial correctness.
Let us select the control points 1, 2, and 3
as depicted in Fig. \ref{flowchart1}
to cut the flowchart into three loop-free paths:
\begin{description}
  \item[path (1..2)] from the starting point 1 to point 2;
  \item[path (2+3)] from point 2 to the final point 3
  via the positive branch;
  \item[path (2--2)] from point 2 to the same point 2
  via the negative branch.
\end{description}
Let us consider all these paths one by one using the following annotations
for the control points:
\begin{enumerate}
\item $Y>1\ \&\ Eps>0$ (i.e. the pre-condition);
\item $Y>1\ \&\ Eps>0\ \&\ \sqrt{Y}\leq X\leq Y$ (the loop invariant);
\item $|X-\sqrt{Y}|\leq Eps$ (i.e. the post-condition).
\end{enumerate}

The first path (1..2) is easy to verify:
{\scriptsize
$$\frac{(Y>1\ \&\ Eps>0)\rightarrow (Y>1\ \&\ Eps>0\ \&\ \sqrt{Y}\leq Y\leq Y)}
{\{Y>1\ \&\ Eps>0\}\ X:=Y\ \{Y>1\ \&\ Eps>0\ \&\ \sqrt{Y}\leq X\leq Y\}}. $$}

The second path (2+3) is not so easy. Let us introduce
a test program construct $\phi?$ as a short-hand for $if\ \phi\ then\ stop\ else\ abort$.
Then verification of the path (after some simplification) is as follows:
{\scriptsize
\begin{equation}\label{proof2+3}
\begin{array}{c}
   (Y>1\ \&\ Eps>0\ \&\ \sqrt{Y}\leq X\leq Y\ \&\ |\frac{Y-X^2}{2X}|< Eps/2)\ \rightarrow\
   |X - \sqrt{Y}|<Eps \\
   \\
  \hline
  \\
   \{Y>1\ \&\ Eps>0\ \&\ \sqrt{Y}\leq X\leq Y\}\
  D:=\frac{Y-X^2}{2X} \ \{|D|< Eps/2\ \rightarrow\
  |X - \sqrt{Y}|<Eps/2\} \\
  \\
  \hline
  \\
  \{Y>1\ \&\ Eps>0\ \&\ \sqrt{Y}\leq X\leq Y\}\
  D:=\frac{Y-X^2}{2X}\ ; \
  |D|< Eps/2?\ \{|X - \sqrt{Y}|<Eps\}
\end{array}
\end{equation}}
The premise
$$(Y>1\ \&\ Eps>0\ \&\ \sqrt{Y}\leq X\leq Y\ \&\ |\frac{Y-X^2}{2X}|< Eps/2)\ \rightarrow\
   |X - \sqrt{Y}|<Eps$$
is valid since in this case we have
$$|X - \sqrt{Y}|=\Big(\frac{|X-\sqrt{Y}|\ (X+\sqrt{Y})}{2X}\Big)\times\Big(\frac{2X}
{X+\sqrt{Y}}\Big) \leq
\frac{Eps}{2}\times\frac{2}{1+\frac{\sqrt{Y}}{X}} < Eps.$$

The proof (also after some simplification) of the third path (2--2) is as follows:
{\scriptsize
\begin{equation}\label{proof2-2}
\begin{array}{c}
\hspace*{10cm} \\
   (Y>1\ \&\ Eps>0\ \&\ \sqrt{Y}\leq X\leq Y\ \&\ |\frac{Y-X^2}{2X}|\geq Eps/2)\ \rightarrow\  \hspace*{\fill} \\
  \hspace*{\fill}
   \rightarrow\ (Y>1\ \&\ Eps>0\ \&\ \sqrt{Y}\leq \frac{Y+X^2}{2X}\leq Y) \\
   \\
  \hline
  \\
  \hspace*{10cm} \\
   \{Y>1\ \&\ Eps>0\ \&\ \sqrt{Y}\leq X\leq Y\}\  \hspace*{\fill} \\
  \hspace*{\fill}
  D:=\frac{Y-X^2}{2X} \hspace*{\fill} \\
  \hspace*{\fill} \{|D|\geq Eps/2\ \rightarrow\
  (Y>1\ \&\ Eps>0\ \&\ \sqrt{Y}\leq X+D\leq Y)\} \\
  \\
  \hline
  \\
  \hspace*{10cm} \\
  \{Y>1\ \&\ Eps>0\ \&\ \sqrt{Y}\leq X\leq Y\}\  \hspace*{\fill} \\
  \hspace*{\fill}
  D:=\frac{Y-X^2}{2X}\ ;  |D|\geq Eps/2?\ ;\ X:=X+D\  \hspace*{\fill} \\
  \hspace*{\fill} \{Y>1\ \&\ Eps>0\ \&\ \sqrt{Y}\leq X\leq Y\}
\end{array}
\end{equation}}
A hint to prove the premise of this derivation:
\begin{itemize}
\item $(Y>1\ \&\ \sqrt{Y}\leq X)\ \rightarrow\ (Y>1\ \&\ \sqrt{Y}\leq \frac{Y+X^2}{2X})$\\
since $(Y>1\ \&\ \sqrt{Y}\leq X)$ implies $X>1$ and, hence, both sides of the AM-GM inequality
 $X\sqrt{Y}\leq \frac{Y+X^2}{2}$ may be divided by $X$;
\item $(Y>1\ \&\ \sqrt{Y}\leq X\leq Y)\ \rightarrow\ (Y>1\ \&\ \frac{Y+X^2}{2X}< Y)$\\
since $(Y>1\ \&\ \sqrt{Y}\leq X\leq Y)$ implies $\frac{Y-X^2}{2X}\leq 0$ and, hence,
$\frac{Y+X^2}{2X}\ =\ X + \frac{Y-X^2}{2X}\leq\ Y + \frac{Y-X^2}{2X}\ \leq\ Y$.
\end{itemize}

\subsection{Termination}\label{TermSQR}
Let us prove below that the loop invariant
$Y>1\ \&\ Eps>0\ \&\ \sqrt{Y}\leq X\leq Y$
implies that every loop iteration reduces the absolute value of $D$ twice at least.

For it let us fix some $y>1$  as the initial value of the variable $Y$,
$\varepsilon>0$ as the initial value of the variable $Eps$,
let $x_1$, $x_2$, $\dots$ $x_n$, $x_{(n+1)}$, $\dots$
be the values of the variable $X$ immediately \emph{before} $1^{st}$, $2^{nd}$,
$\dots$ $n$-th, $(n+1)$-th, etc., iteration of the loop
for this fixed initial value $y$ of $Y$,
and let $d_1$, $d_2$, $\dots$ $d_n$, $d_{(n+1)}$, $\dots$ be the values of the variable $D$
immediately \emph{after} $1^{st}$, $2^{nd}$, $\dots$ $n$-th, $(n+1)$-th, etc.,
iteration of the loop (also for the same fixed initial value $y$ of $Y$).
In particular, $x_1=y$ and
$d_n=\frac{y-x_n^2}{2x_n}$, $x_{(n+1)}=x_n+d_n$ for all $n>0$.

Let us express $d_{(n+1)}$ in terms of $d_n$:
\begin{center}
$d_{(n+1)} = \frac{y-x_{(n+1)}^2}{2x_{(n+1)}} =
\frac{y-(x_{n}+d_n)^2}{2(x_{n}+d_n)} =
\frac{y-(\frac{y+x_n^2}{2x_n})^2}{2\frac{y+x_n^2}{2x_n}} =$ \hspace*{\fill} \\
\hspace*{\fill} $= -\frac{(y-x_n^2)^2 x_n}{4x_n^2 (y+x_n^2)} =
-\frac{d_n^2 x_n}{y+x_n^2} = - \frac{d_n^2}{2x_{(n+1)}}$.
\end{center}
Note that all values $d_1$, $d_2$, $\dots$ $d_n$, $d_{(n+1)}$, $\dots$
are negative due to the loop invariant.
Hence
$$
\frac{|d_{(n+1)}|}{|d_n|} = \frac{d_{(n+1)}}{d_n} = - \frac{d_n}{2x_{(n+1)}} =
\frac{1}{2}\times\frac{x_n^2 - y}{x_n^2 +y} < \frac{1}{2}.
$$
It implies $|d_{(n+1)}| < \frac{y - \sqrt{y}}{2^n}$, i.e. the algorithm terminates
after at most $$1+\log_2\frac{y - \sqrt{y}}{\varepsilon}$$ iterations of the loop.

\section{Towards machine-oriented square root algorithm}\label{TowMachSqrt}
\subsection{Improved square root algorithms based on until-loop}\label{impSQRT}
\begin{figure}[t]
\centering
  \includegraphics[width=8cm]{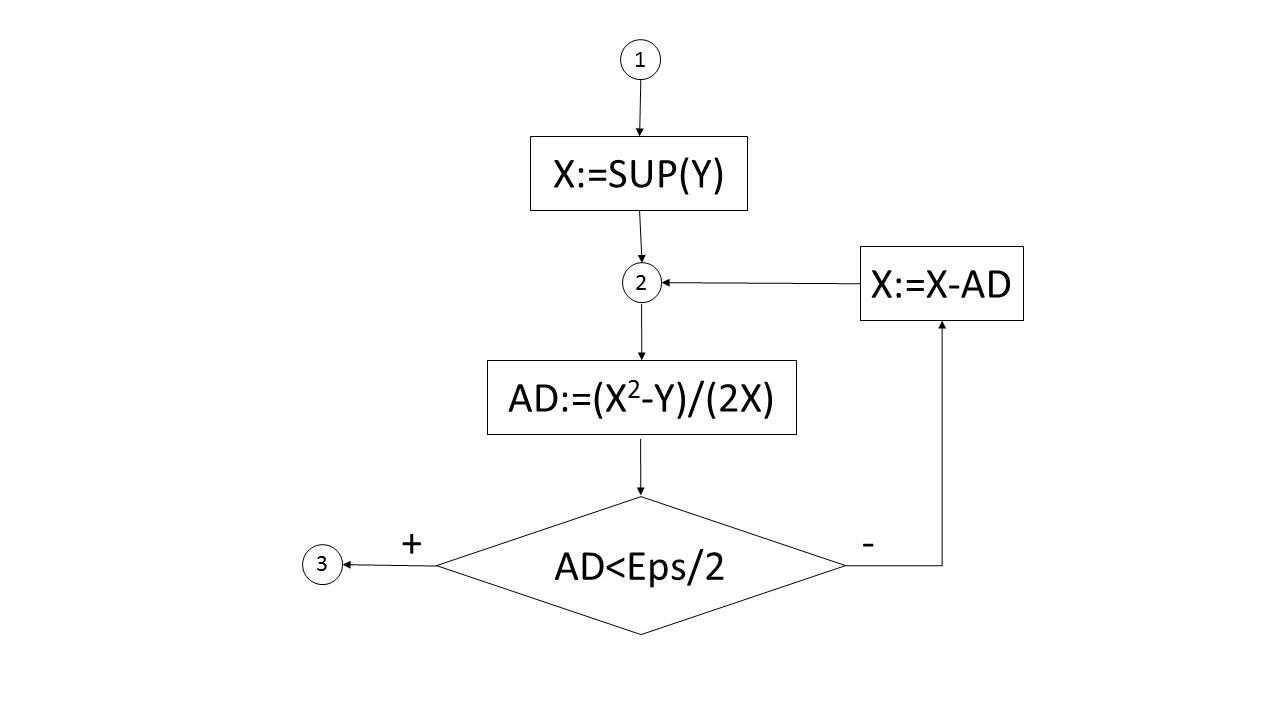}\\
  \caption{A flowchart of the improved (optimized) algorithm $ISQR$}
  \label{flowchart2}
\end{figure}

In spite being very efficient (due to a logarithmic complexity) the algorithm may be improved (optimized).
Firstly, since we study case when $1<Y$ and know (from the loop invariant) that $\sqrt{Y}\leq X\leq Y$,
it makes sense to compute directly the absolute value $AD:=\frac{X^2-Y}{2X}$ of $D$
instead of computing $D:=\frac{Y-X^2}{2X}$ and then $|D|$ in the loop condition.
Next, we may use a \emph{fast} hash function $SUP:(1,\infty)\rightarrow (1,\infty)$
to compute \emph{good}  initial upper approximations
instead of a very rough initial upper approximation used in the algorithm $SQR$.
(For example, it may be rounded-up square roots.)
While the first optimization just saves on each loop iteration on calls of the function computing the absolute value,
the second one reduces the number of loop iterations.
Fig. \ref{flowchart2}  shows a flowchart of the improved algorithm that we refer as the algorithm $ISQR$ in the sequel.

For example if the function $SUP$ returns the rounded-up square roots,
$y>1$ is the initial (input) value of the variable $Y$,
and $\varepsilon>0$ is the initial (input) value of variable $Eps$ (accuracy)
then $0\leq SUP(y) - \sqrt{y}<1$ and, hence, an upper bound for the number of the loop iterations in the algorithm $ISQR$ is
$$1-\log_2\varepsilon\ \geq\ 1+\log_2\frac{SUP(y) - \sqrt{y}}{\varepsilon}$$
instead of an upper bound $$1 + \log_2y - \log_2\varepsilon\ \geq\ 1+\log_2\frac{y - \sqrt{y}}{\varepsilon}$$
for the number of the loop iterations in the non-optimized algorithm $SQR$.

To prove the following total correctness assertion
\begin{center}
$[Y>1\ \&\ Eps>0\ \&\ \forall y\in (1,+\infty): \sqrt{y}\leq SUP(y) \leq y)]$ \hspace*{\fill} \\
\hspace*{\fill} $ISQR\ [|X - \sqrt{Y}|<Eps]$
\end{center}
we need to prove the partial correctness only since the termination are proved already
by providing an upper bound for the number of the loop iterations.

For proving the partial correctness we may use the same control points 1, 2, and 3
to cut the flowchart into three loop-free paths
(1..2), (2+3), (2--2), and the same annotations for the control points
2 (the loop invariant) and 3 (the post-condition) as for the algorithm $SQR$,
but need to extend the precondition ($Y>1\ \&\ Eps>0$) of the $SQR$
by specification of the function $SUP$:
$$\forall y\in (1,+\infty): \sqrt{y}\leq SUP(y) \leq y.$$
The above proof (\ref{proof2+3}) of the path (2+3) and the proof (\ref{proof2-2}) of the path (2--2) remains valid.
The first path (1..2) is easy to verify:
{\scriptsize
\begin{equation}\label{proof1.2}
\begin{array}{c}
\hspace*{10cm} \\
 (Y>1\ \&\ Eps>0\ \&\ \forall y\in (1,+\infty): \sqrt{y}\leq SUP(y) \leq y)\ \rightarrow\  \hspace*{\fill} \\
  \hspace*{\fill}
   \rightarrow\ (Y>1\ \&\ Eps>0\ \&\ \sqrt{Y}\leq SUP(Y)\leq Y) \\
   \\
  \hline
  \\
 \{Y>1\ \&\ Eps>0\ \&\ \forall y\in (1,+\infty): \sqrt{y}\leq SUP(y) \leq y)\} \hspace*{\fill} \\
 \hspace*{\fill} X:=SUP(Y)\  \{Y>1\ \&\ Eps>0\ \&\ \sqrt{Y}\leq X\leq Y\}
\end{array}
\end{equation}}

\subsection{For-loop-based square root algorithms\\ (when \emph{more} means \emph{better}) }\label{forSQRT}
\begin{figure}[t]
\centering
  \includegraphics[width=8cm]{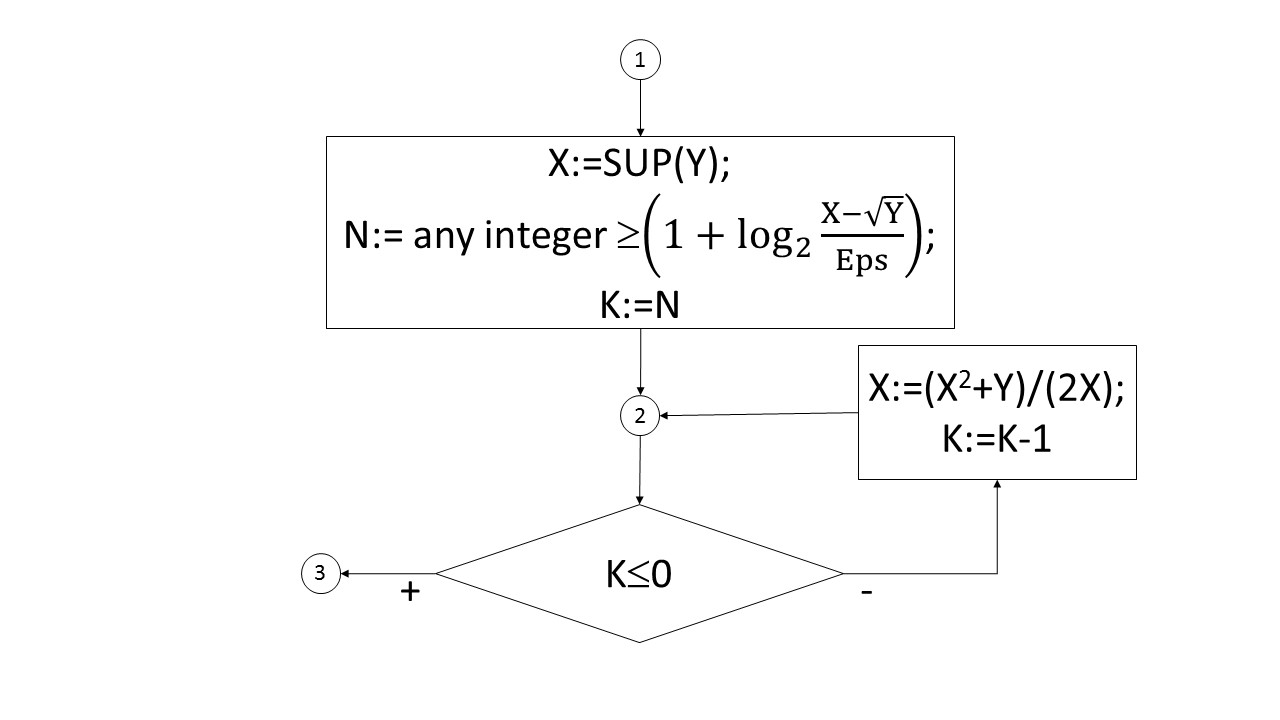}\\
  \caption{A flowchart of the non-adaptive for-loop-based algorithm $FSQR$}
  \label{flowchart3}
\end{figure}
As we already proved,
for every initial value $y>1$ of the variable $Y$ and every initial value $\varepsilon>0$ of the variable $Eps$
termination of the improved square root algorithm is guaranteed
after (at most)  $1+\lceil\log_2\frac{SUP(y) - \sqrt{y}}{\varepsilon}\rceil$ loop iterations,
where $\lceil\dots\rceil$ is integer round-up function.
Hence is possible to compute approximations for the square root by
a non-adaptive for-loop-based algorithm $FSQR$
which flowchart depicted in Fig. \ref{flowchart3}.
The algorithm uses a non-deterministic assignment
\begin{equation}\label{AssInF}
N:= \mbox{ any integer } \geq \big(1+ \log_2\frac{X-\sqrt{Y}}{Eps}\big)
\end{equation}
(where $\lceil\dots\rceil$ stays for integer rounding up).
The corresponding correctness assertion is
\begin{equation}\label{FSQRass}
\begin{array}{c}
[Y>1\ \&\ Eps>0\ \&\ \forall y\in (1,+\infty): \sqrt{y}\leq SUP(y) \leq y)] \hspace*{2cm} \\
\hspace*{\fill} FSQR\ [|X - \sqrt{Y}|<\frac{Eps}{2}].
\end{array}
\end{equation}

Termination of the algorithm $FSQR$ is guaranteed by design since it is for-loop-based.
Informally speaking the partial correctness of the algorithm follows
from the partial correctness of the algorithm $ISQR$:
while $\frac{X^2-Y}{2X}\geq Eps$ values of $X$ in both algorithms are equal in each iteration,
and then $FSQR$ exercises several more iterations that move value of $X$ closer to $\sqrt{Y}$.
Nevertheless we would like to make this argument more formal and in Floyd-Hoare style \cite{Gries81}.

Let us select the control points 1, 2, and 3
as depicted in Fig. \ref{flowchart3}
and annotate them as follows:
\begin{enumerate}
\item $Y>1\ \&\ Eps>0\ \&\ \forall y\in (1,+\infty): \sqrt{y}\leq SUP(y) \leq y)$;
\item $Y>1\ \&\ Eps>0\ \&\ \sqrt{Y}\leq X \leq Y\ \&\ N\geq 1+\log_2\frac{SUP(Y) - \sqrt{Y}}{Eps}\ \&$ \\
\hspace*{\fill} $\&\ 0\leq K\leq N\ \&\ X-\sqrt{Y}\leq \frac{SUP(Y) - \sqrt{Y}}{2^{N-K}}$;
\item $|X-\sqrt{Y}|\leq \frac{Eps}{2}$ (i.e. the post-condition).
\end{enumerate}

Proof of the path (1..2) is trivial since at the end of this path
$N\geq 1+\lceil\log_2\frac{SUP(Y) - \sqrt{Y}}{Eps}\rceil$, $N-K=0$ and $X=SUP(Y)$.

Proof of the path (2--2) just follows the proof of the similar path for the algorithm $SQR$
with the following addendum:
$0\leq K\leq N$ before the loop implies $0\leq K\leq N$ after the loop because of
the loop condition $K>0$ that holds on this path before the assignment $K:=K-1$.

Proof of the path (2+3) is more complicated:
at the end of the path
\begin{enumerate}
\item \label{one} $K=0$ (due to the invariant at start of the path and the loop condition);
\item \label{two} $\sqrt{Y}\leq X$ (due to the invariant at start of the path);
\item \label{three} according to (\ref{two}) $|X-\sqrt(Y)|=X-\sqrt{Y}$;
\item \label{four} according to (\ref{one}), (\ref{three}) and the invariant
      $|X-\sqrt(Y)|\leq \frac{SUP(Y) - \sqrt{Y}}{2^N}$;
\item \label{five} $\frac{SUP(Y) - \sqrt{Y}}{2^N}\leq
    \frac{SUP(Y) - \sqrt{Y}}{2^{1+\log_2\frac{SUP(Y) - \sqrt{Y}}{Eps}}}= \frac{SUP(Y) - \sqrt{Y}}{2\times\frac{SUP(Y) - \sqrt{Y}}{Eps}} =
    \frac{Eps}{2}$\\ (because $N\geq 1+\log_2\frac{SUP(Y) - \sqrt{Y}}{Eps}$);
\item \label{six} according to (\ref{four}) and (\ref{five})
$|X-\sqrt{Y}|\leq \frac{Eps}{2}$.
\end{enumerate}

We have proved  a stronger assertion for $FSQR$ than for $SQR$ and $ISQR$.
Moreover, the proof implies that \emph{more} iterations means \emph{better} accuracy of computations
(in the precise arithmetic, of course).

\section{Square root algorithm for fix-point arithmetics}\label{fixSQR}
\subsection{Fix-point machine arithmetics}\label{FixPointArith}
One of the problems with the improved and for-loop-based algorithms is how to implement an efficient function $SUP$.
A hint is use of a numeric data type $T$ with a (huge maybe) finite set of values $Val_T\subset \mathbb{R}$
instead of an infinite set $\mathbb{R}$.
Then the function $SUP$ may be implemented in two steps:
\begin{itemize}
\item define an efficient rounding up function $round:Val_T\rightarrow Val_T$,
\item pre-compute and memorize a look-up table $root$ with good upper approximations for the roots for each of the rounded values.
\end{itemize}
Further details and steps depend on selected numeric data type.
In this and the next sections we study \emph{fix-point} numeric data and algorithms with \emph{fix-point} arithmetic.
(We study of $floating-point$ data and algorithms later in the section \ref{fltSect}.)

We understand fix-point numeric data type $T$ as follows:
\begin{itemize}
\item the set of values $Val_T$ is a finite subset of mathematical reals $\mathbb{R}$
such that
\begin{itemize}
\item it comprises all reals in some finite range $[-\inf_T,\sup_T]$, where $\inf_T>2$, $\sup_T>2$, with some fixed step $\frac{1}{2}>\delta_T>0$,
\item and includes all integer numbers $Int_T$ in this range $[-\inf_T,\sup_T]$;
\end{itemize}
\item legal binary arithmetic operations are
\begin{itemize}
\item addition and subtraction;
if not the range overflow exception then these operations are precise: they equal to the standard mathematical operations
assuming their mathematical  results fall in the range $[-\inf_T,\sup_T]$
(and due to this reason are denoted as $+$ and $-$);
\item multiplication $\otimes$ and division $\oslash$;
these operations are approximate but correctly rounded in the following sense:
for all $x,y\in Val_T$
\begin{itemize}
\item if $x\times y\in Val_T$ then $x\otimes y = x\times y$;
\item if $x/y\in Val_T$ then $x\oslash y = x/y$;
\item if $x\times y\in [-\inf_T,\sup_T]$ then $|x\otimes y - x\times y|<\delta_T/2$;
\item if $x/y\in [-\inf_T,\sup_T]$ then $|x\oslash y - x/y|<\delta_T/2$.
\end{itemize}
\end{itemize}
\item legal binary relations are equality and all standard inequalities;
these relations are precise, i.e. they equal to the standard mathematical relations
(and due to this reason are denoted as $=$, $\neq$, $\leq$, $\geq$, $<$, $>$).
\end{itemize}

Due to the assumptions about the set of values
$$Val_T = \{n\times\delta_T\ :\ n\in \mathbb{Z} \mbox{ and } -\mbox{inf}_T\leq n\times\delta_T\leq\mbox{sup}_T\};$$
according the assumptions about integer values $Int_T$ within the range of $Val_T$
$$[-2..2]\subseteq Val_T \mbox{ and } \frac{1}\delta_T\in\mathbb{N}.$$

In case when multiplication is guaranteed to be precise
(the mathematical product is in $Val_T$)
then let us use the standard notation $\times$ instead of $\otimes$;
similarly in case when  division is guaranteed to be precise
(the mathematical dividend is in $Val_T$)
then let us use the standard notation $/$ instead of $\oslash$.

\subsection{Fix-point variant of the square root: algorithm and specification}\label{SpecfixSQRT}
\begin{figure}[t]
\centering
  \includegraphics[width=8cm]{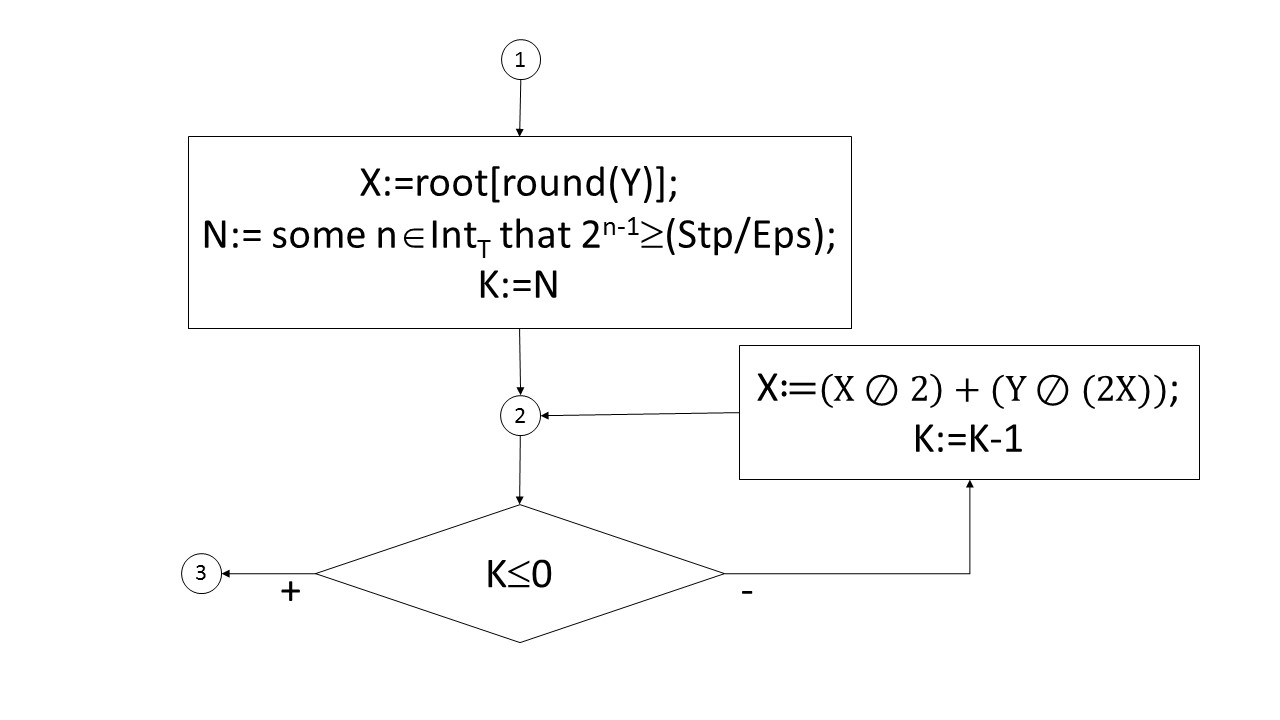}\\
  \caption{A flowchart of the square root algorithm $fixSQR$ for fix-point arithmetic}
  \label{flowchart4}
\end{figure}
A non-adaptive algorithm $FSQR$ (Fig. \ref{flowchart3}) that uses mathematical operations transforms into
algorithm $fixSQR$ (Fig. \ref{flowchart4}) that uses machine fix-point operations.
This algorithm also (as $FSQR$) uses a non-deterministic assignment operator
\begin{equation}\label{AssInFix}
N:= \mbox{ some } n\in Int_T \mbox{ that } 2^{n-1}\geq\frac{Stp}{Eps}
\end{equation}
that differs from the assignment (\ref{AssInF}) by use of \emph{some} instead of \emph{any}:
this difference means that later we \emph{select} the value instead of use an \emph{arbitrary} one.

In the new algorithm we use an additional variable $Stp$ for a positive value in $Val_T$, an array $root$, and a function $round$
that have the following properties:
\begin{description}
\item[STEP:] value of $Stp$  is a multiple of the accuracy $Eps$, divides $\sup_T$
and is used to define
the set $Arg_{Stp} = \{n\times Stp\ :\ n\in\mathbb{N} \mbox{, and } 1<n\times stp\leq \sup_T\}$;
\item[ROOT:] $root$ is a pre-computed look-up table indexed by $Arg_{Stp}$ such that
$root[v]-\delta_T<\sqrt{v}\leq root[v]$ for each index $v\in Arg_{Stp}$;
\item[ROUND:] the function $round:Val_T\rightarrow Arg_{Stp}$ is a rounding-up such that
$round(u)-step<u\leq round(u)$ for each $u\in Val_T$, $u>1$.
\end{description}
\textbf{Comment on the STEP property:} we consider as a very natural the
assumption that
\begin{itemize}
\item $Stp$ is a multiple of the accuracy $Eps$ since in the ``limit'' case $Eps=\delta_T$ and this $Eps$ divides any $Stp\in Val_T$;
\item $Stp$ divides the ``extreme'' value  $\sup_T$ because this value should be provided with a pre-computed square root upper approximation.
\end{itemize}

We are ready to specify correctness of the square root algorithm $fixSQR$ with fix-point arithmetic:
\begin{equation}\label{fixSQRass}
\begin{array}{c}
[Y\in Val_T\ \&\ Y>1\ \&\ Eps\in Val_T\ \&\ Eps>0\ \& \hspace*{3cm} \\
\hspace*{\fill} \&\ \mathbf{STEP}\ \&\ \mathbf{ROOT}\ \&\ \mathbf{ROUND}]\hspace*{2.5cm} \\
\hspace*{\fill} fixSQR\ [|X - \sqrt{Y}|<\big(\frac{Eps}{2}\ +\ N\times\delta_T\big)].
\end{array}
\end{equation}

\section{Pen-and-paper verification of $fixSQR$\\
(\emph{more} may be \emph{worse})}\label{verFixSQR}
Termination of the algorithm $fixSQR$ is straightforward since it is a for-loop-based algorithm.
So we need to prove partial correctness only.
We do this proof below by \emph{adjustment} (or comparison) of runs of
algorithm $fixSQR$ with fix-point arithmetics and algorithm $FSQR$ with precise arithmetics.

Let us select and fix hereafter initial values $x$, $y$, $s$, $\varepsilon$ for the variables $X$, $Y$, $Stp$, and $Eps$,
and a look-up table $root$ and a function $round$ such that meet the precondition in (\ref{fixSQRass}).
Let $SUP:(1,\infty)\rightarrow (1,\infty)$ be the function defined as follows:
$$SUP(u) = \left\{\begin{array}{l}
root[v] \mbox{, if } u\in (1,\mbox{sup}_T] \mbox{, } v\in Arg_{Stp} \mbox{ and } u\in (v-s,v] \mbox{;}\\
u \mbox{ otherwise.}  \end{array}\right.$$
Then this function and the initial values of $X$, $Y$, and $Eps$ meet the precondition in (\ref{FSQRass}).

Let $nn$ be a particular value assigned to $N$ by the non-deterministic assignment operator (\ref{AssInFix})
in the algorithm $fixSQR$.
Remark that $SUP(y) - \sqrt{y}\leq s$ due to the following arguments:
$$\big(SUP(y)\big)^2 \leq y + s \leq y + 2s\sqrt{y} + s^2 = (\sqrt{y} + s)^2;$$
hence this value $nn$ is also a legal value of the non-deterministic expression
$$\mbox{ any integer } \geq \big(1+ \log_2\frac{SUP(y)-\sqrt{y}}{\varepsilon}\big)$$
that is the right-hand expression in the assignment (\ref{AssInF}).
It implies that both algorithms $FSQR$ and $fixSQR$ have legal runs with the
initial values $x$, $y$, $\varepsilon$ for the variables $X$, $Y$, and $Eps$
where both have exactly $nn$ iterations of their loops.

Let  $x^{\prime}_{0}$, $\dots$ $x^{\prime}_{nn}$ and $x^{\prime\prime}_{0}$, $\dots$ $x^{\prime\prime}_{nn}$
values of the variable $X$ in these runs \emph{after} $0$-iterations, $\dots$ $nn$-iterations of the corresponding loop.
(In particular, $x=x^{\prime}_{0}=x^{\prime\prime}_{0}$ is the initial value of $X$ and
$x^{\prime}_{nn}$ and $x^{\prime\prime}_{nn}$ are the final values of the variable upon termination.)

Let us prove by induction on $k\in[0..nn]$ that
\begin{equation}\label{xprimes}
  |x^{\prime}_{k} - x^{\prime\prime}_{k}|\leq k\delta_T.
\end{equation}
\begin{description}
  \item[Basis:] $x=x^{\prime}_{0}=x^{\prime\prime}_{0}$.
  \item[Assumption:]  $|x^{\prime}_{k} - x^{\prime\prime}_{k}|\leq k\delta_T$ for all $k\in[0..n]$, where $n<nn$.
  \item[Step:] Let $\Delta = x^{\prime}_{n} - x^{\prime\prime}_{n}$; ($|\Delta|\leq n\delta_T$ due to the assumption.)
  \begin{enumerate}
  \item $x^{\prime}_{n+1} = \frac{x^{\prime}_{n}}{2} + \frac{y}{2x^{\prime}_{n}}$;
  \item $x^{\prime\prime}_{n+1} = x^{\prime\prime}_{n}\oslash 2 + y\oslash(2x^{\prime\prime}_{n}) =
  \big(\frac{x^{\prime\prime}_{n}}{2} + \delta_a\big) + \big(\frac{y}{2x^{\prime\prime}_{n}} + \delta_b\big)$
  where $|\delta_a|,|\delta_b|\leq \frac{\delta_T}{2}$;
  \item \label{Dif} $|x^{\prime\prime}_{n+1} - x^{\prime}_{n+1}| = |\frac{x^{\prime\prime}_{n} - x^{\prime}_{n}}{2} +
  \frac{y}{2}\big(\frac{1}{x^{\prime\prime}_{n}} - \frac{1}{x^{\prime}_{n}}\big) + (\delta_a+\delta_b)| \leq$\\
  \hspace*{\fill} $\leq \frac{|x^{\prime\prime}_{n} - x^{\prime}_{n}|}{2} +
  \frac{y}{2}\big|\frac{1}{x^{\prime\prime}_{n}} - \frac{1}{x^{\prime}_{n}}\big| + |\delta_a+\delta_b| \leq
  \frac{|\Delta|}{2} +
  \frac{y}{2}\big|\frac{1}{x^{\prime\prime}_{n}} - \frac{1}{x^{\prime}_{n}}\big| + \delta_T$;
  \item \label{Delta}(Taylor' expansion) $\frac{1}{x^{\prime\prime}_{n}} = \frac{1}{x^{\prime}_{n}+\Delta} =
  \frac{1}{x^{\prime}_{n}} - \frac{\Delta}{(x^{\prime}_{n})^2} + \theta$,\\
  \hspace*{\fill} where $\theta$ is a converging alternating series,
  $|\theta - \frac{\Delta}{(x^{\prime}_{n})^2}|\leq |\frac{\Delta}{(x^{\prime}_{n})^2}|$;
  \item \label{dif} (from \ref{Delta}) $\frac{y}{2}\big|\frac{1}{x^{\prime\prime}_{n}} - \frac{1}{x^{\prime}_{n}}\big| =
  \frac{y}{2}\big|\theta - \frac{\Delta}{(x^{\prime}_{n})^2}\big| \leq \frac{y}{2}\times \frac{|\Delta|}{(x^{\prime}_{n})^2} \leq
  \frac{y}{2}\times \frac{|\Delta|}{y} = \frac{|\Delta|}{2}$;
  \item \label{DIF} (from \ref{Dif} and \ref{dif})
  $|x^{\prime\prime}_{n+1} - x^{\prime}_{n+1}| \leq \frac{|\Delta|}{2} +
  \frac{y}{2}\big|\frac{1}{x^{\prime\prime}_{n}} - \frac{1}{x^{\prime}_{n}}\big| + \delta_T \leq
  \frac{|\Delta|}{2} + \frac{|\Delta|}{2} + \delta_T$;
  \item (from \ref{DIF}) $|x^{\prime\prime}_{n+1} - x^{\prime}_{n+1}| \leq |\Delta| + \delta_T \leq n\delta_T + \delta_T =
  (n+1)\delta_T$.
   \end{enumerate}
\end{description}

According to the the proven total correctness assertion (\ref{FSQRass})
$|x^{\prime}_{nn} - \sqrt{y}|\leq \frac{\varepsilon}{2}$;
together with the proven property (\ref{xprimes}) it implies
$|x^{\prime\prime}_{nn} - \sqrt{y}|\leq \big(\frac{\varepsilon}{2} + nn\times \delta\big)$;
since $x^{\prime\prime}_{nn}$ and $nn$ are the values of the variables $X$ and $N$ in the algorithm $fixSQR$,
it finishes the proof of the assertion (\ref{fixSQRass}).

One can remark that correctness of the assertion (\ref{fixSQRass}) implies that \emph{more} iterations
of the loop may be \emph{worse} in accuracy (due to the addend $N\times\delta_T$ in the postcondition).

\begin{figure}[t]
  \centering
  \includegraphics[width=8cm]{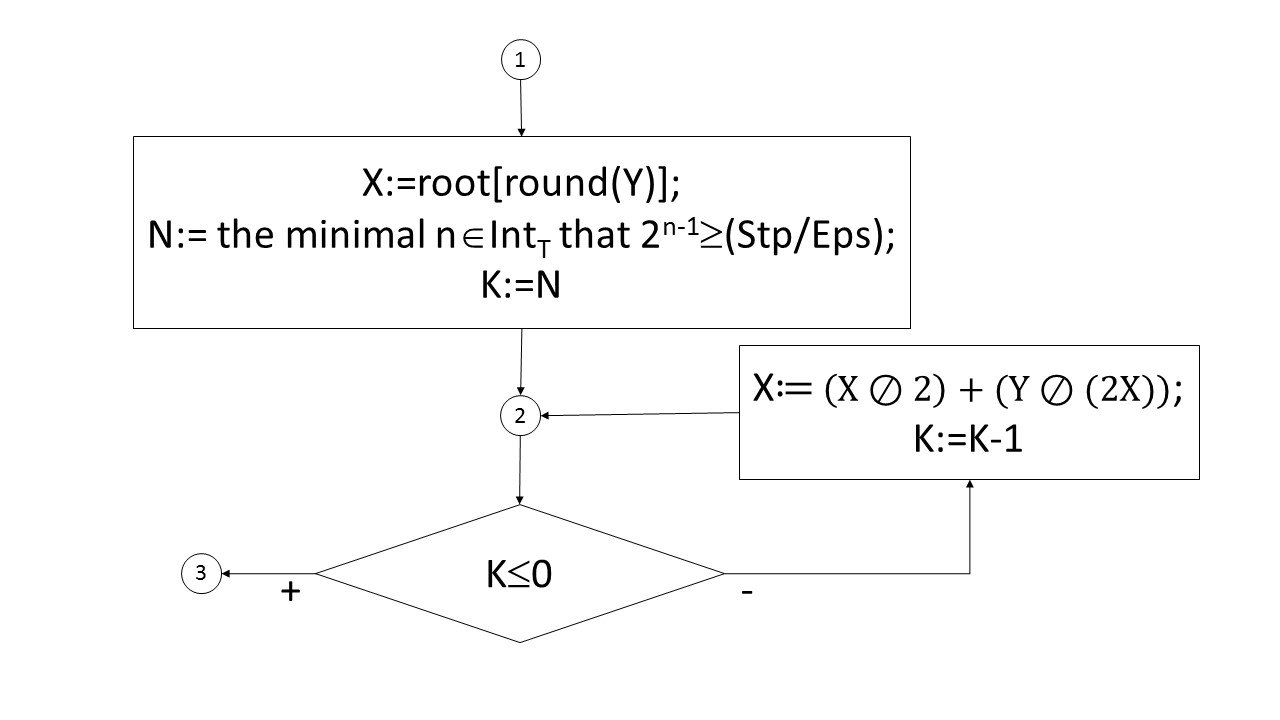}\\
  \caption{A flowchart of the square root algorithm $mixSQR$ for fix-point arithmetic}\label{flowchart5}
\end{figure}

Our proof of the assertion (\ref{fixSQRass}) implies correctness of the following assertion
\begin{equation}\label{mufixSQRass}
\begin{array}{c}
[Y\in Val_T\ \&\ Y>1\ \&\ Eps\in Val_T\ \&\ Eps>0\ \&\ \hspace*{3cm} \\
\&\ EPS\geq 2\delta_T(2+\log_2\frac{Stp}{Eps})\ \& \hspace*{\fill} \\
\hspace*{\fill} \&\ \mathbf{STEP}\ \&\ \mathbf{ROOT}\ \&\ \mathbf{ROUND}]\hspace*{2.5cm} \\
\hspace*{\fill} mixSQR\ [|X - \sqrt{Y}|<Eps].
\end{array}
\end{equation}
where $mixSQR$ is algorithm depicted on Fig. \ref{flowchart5}.
The assertion is valid due to the arguments represented in the next paragraph.

The algorithm $mixSQR$ is a specialization of the algorithm $fixSQR$
with deterministic assignment
$$N:= \mbox{ the minimal } n\in Int_T \mbox{ that } 2^{n-1}\geq\frac{Stp}{Eps}$$
instead of the nondeterministic assignment (\ref{AssInFix}).
The precondition in (\ref{mufixSQRass}) expands the precondition in (\ref{fixSQRass}) by the
addend $EPS\geq 2\delta_T(2+\log_2\frac{Stp}{Eps})$
that means that the interval $[(1+\log_2\frac{Stp}{Eps}), \frac{Eps}{2\delta_T}]$ has length $\geq 1$,
i.e. contains an integer.
Let $y$, $\varepsilon$, $s$ be initial values of $Y$, $Eps$, and $Stp$ that satisfy the precondition in (\ref{mufixSQRass})
(and hence the precondition in (\ref{fixSQRass})), and
let $nn$ be  the minimal $n\in Int_T$ that $2^{n-1}\geq\frac{Stp}{Eps}$ (i.e. the value assigned to the variable $N$).
Since $y$, $\varepsilon$, $s$ satisfy the precondition in (\ref{fixSQRass}), the algorithm $mixSQR$ stops on these initial data
with final values of the variables that satisfy the postcondition in (\ref{fixSQRass}), i.e.
$|x^\prime - \sqrt{y}|<\big(\frac{\varepsilon}{2}\ +\ nn\times \delta_T\big)$, where $x^\prime$ is the final value of $X$;
since the value of $N$ is $nn$ then $nn\times \delta_T<\frac{\varepsilon}{2\delta_T}\times \delta_T = \frac{\varepsilon}{2}$;
put it altogether we get that $|x^\prime - \sqrt{y}|<\varepsilon$, i.e. the postcondition in (\ref{mufixSQRass}) is true.

\section{Square root algorithm for floating-point arithmetic}\label{fltSect}
In contrast to the fix-point numeric data type in the subsection \ref{FixPointArith},
we aren't going to specify properties of floating-point arithmetic operations
(since we don't need them to compute square root function) but
just the properties of  the set of floating-point values and couple of type-casting operations
(that convert floating-point values into fix-point values and back).

Let $T$ be a fix-point numeric data type that satisfies the properties specified in the subsection \ref{FixPointArith}.
We understand floating-point numeric data type $F$ as follows:
\begin{itemize}
\item the set of values $Val_F$ is a finite subset of mathematical reals $\mathbb{R}$
that comprises some reals in some finite range $[-\inf_F,\sup_F]$, where $\inf_F,\sup_F>2$ and $\{0,\inf_F,\sup_F\}\subseteq Val_F$;
\item there are two unary operations $Man:Val_F\rightarrow Val_T$ (called \emph{mantissa}),
$Exp:Val_F\rightarrow Int_T$ (called \emph{exponent}), and an integer constant $\beta_T\in Int_F$
(called \emph{exponent base} or just \emph{base})
such that for all positive $x\in Val_F$
\begin{itemize}
\item $1< Man(x)< \frac{sup_T}{\beta_F}$;
\item if $inf_T$ is odd then $-inf_T< Exp(x)$ else $-inf_T\leq Exp(x)$;
\item $x= Man(x)\times \beta_F^{Exp(x)}$.
\end{itemize}
\end{itemize}

Firstly remark that according to our definition of mantissa, it ranges in $(1,\frac{sup_T}{\beta_F})$
while the most common definition says that the mantissa ranges in $[0.1,1)$.
We adopt the above definition as a variation and of the standard one due to the following reasons:
the right end of the range ($\frac{sup_T}{\beta_F}$) is parameterised by parameters that characterise
numeric types $T$ and $F$ and hence is more general than any fixed right end;
the left end $1$ of the range is excluded because we want to use a verified algorithm $mixSQR$
to compute in fix-point arithmetic an approximation of the square root form the mantissa.

Next remark that in the property $x= Man(x)\times \beta_F^{Exp(x)}$
we use $\times$ for the precise mathematical multiplication
and assume that the right-hand side product is exactly computable on computer.
This assumption is based on a conventional representation of a floating-point value in the computer memory
as a pair consisting of mantissa and exponent with opportunity to extract the mantissa
and the exponent separately and precisely (we use operations $Man$ and $Exp$)
and then reconstruct the value back (and save it in the memory)
by coupling the mantissa and the exponent (we represent it by using mathematical multiplication $\times$).

\begin{figure}[t]
  \centering
  \includegraphics[width=8cm]{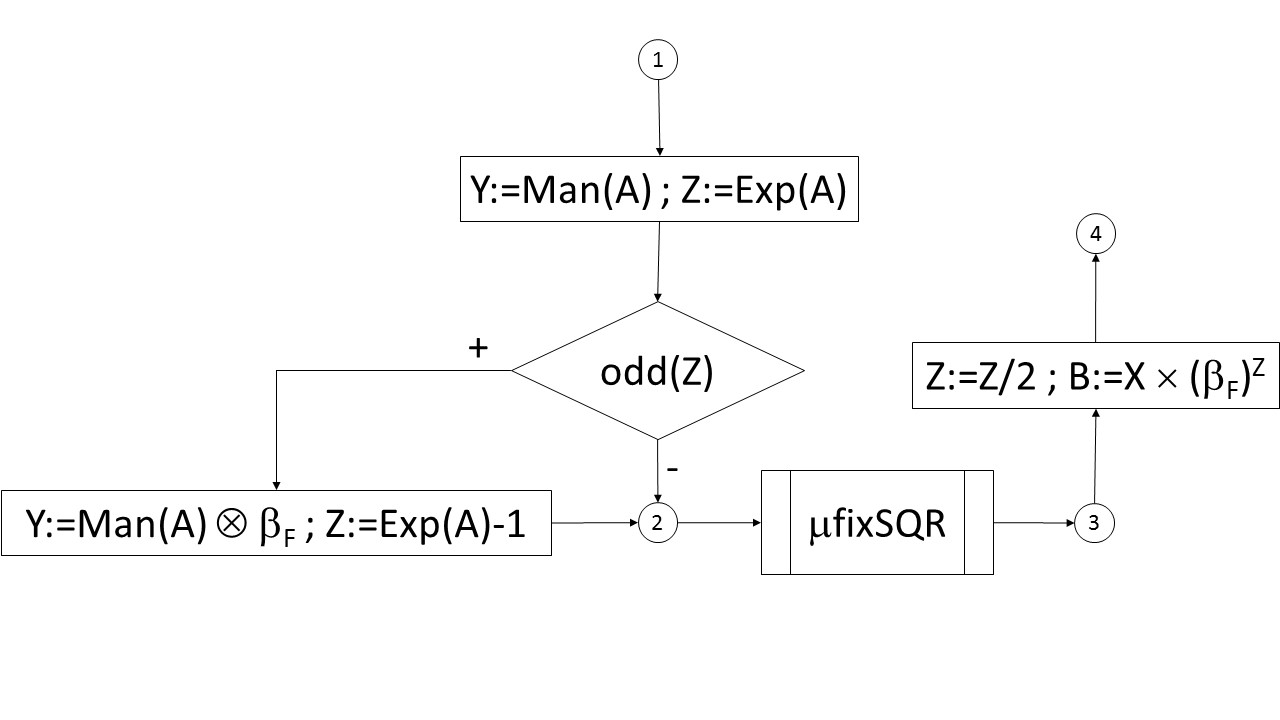}\\
  \caption{A flowchart of the square root algorithm $fltSQR$ for floating-point arithmetic}\label{flowchart6}
\end{figure}

The algorithm $fltSQR$ to compute floating-point approximations of the square root function
 for floating-point argument is presented in Fig. \ref{flowchart6}.
In this algorithm
\begin{itemize}
\item $mixSQR$ is the algorithm  from Fig. \ref{flowchart5},
\item an ``input'' variable $A$ and the ``output'' variable $B$ are of the floating-point type $F$,
\item another ``input'' variable $Eps$ has the fix-point type $T$,
\item a variable $Z$ is of the fix-point type $T$ (but range within integers $Int_T$),
\item a machine operation $\otimes$ is the fix-point multiplication (specified in the subsection \ref{FixPointArith}),
\item and, finally, a constant $\beta_F$ is the exponent base (i.e. a fixed integer of type $T$).
\end{itemize}
Recall that the  algorithm $mixSQR$ uses (within its scope) its own ``local''  variables and a constant:
\begin{itemize}
\item the ``output'' and ``input'' variables $X$ and $Y$ are of the fix-point type $T$,
\item the variables $K$ and $N$ are also of the fix-point type $T$
(but range within integers $Int_T$),
\item the variable $Stp$ of the fix-point type $T$ is the step of indexes of the look-up table
$root$;
\item the look-up table $root$ is an array of the fix-point type $T$ contains pre-computed upper approximations for square root for indexes;
\item and a constant $\delta_T$ is the step of the fix-point type $T$
(i.e. a fixed real value of type $T$ that is the minimal positive value of this type).
\end{itemize}

Specification of the algorithm $fltSQR$ follows below:
\begin{equation}\label{fltSQRass}
\begin{array}{c}
[A\in Val_F\ \&\ Y>0\ \&\ Eps\in Val_T\ \&\ Eps>0\ \&\ \hspace*{3cm} \\
\hspace*{2ex} \&\ EPS\geq 2\delta_T(2+\log_2\frac{Stp}{Eps})\ \& \hspace*{\fill} \\
\hspace*{\fill} \&\ \mathbf{STEP}\ \&\ \mathbf{ROOT}\ \&\ \mathbf{ROUND}]\hspace*{2.5cm} \\
\hspace*{\fill} fltSQR\ [|B - \sqrt{A}|<\big(Eps+\frac{\delta_T}{2\sqrt{\beta_F}}\big)\times \beta^{\lfloor\frac{Exp(A)}{2}\rfloor}],
\end{array}
\end{equation}
where $\lfloor\dots\rfloor$ is integer round-down function.

The assertion is easy to verify since the algorithm $mixSQR$ is verified already and the algorithm $fltSQR$ has loop-free flowchart
(since the only loop is hidden inside $mixSQR$ in this chart).
The only thing we need are annotations for control points on the chart:
\begin{enumerate}
\item the precondition from (\ref{fltSQRass});
\item \label{YZ} the precondition from (\ref{mufixSQRass}) extended by the following two conjuncts:
\begin{enumerate}
\item \label{Y} $Y=\ if\ odd(Exp(A))\ then\ Man(A)\otimes\beta_F\ else\ Man(A)$;
\item \label{Z} $Z=\ if\ odd(Exp(A))\ then\ (Exp(A)-1)\ else\ Exp(A)$;
\end{enumerate}
\item the postcondition from (\ref{mufixSQRass}) extended by the same two conjuncts \ref{Y} and \ref{Z};
\item the postcondition from (\ref{fltSQRass}).
\end{enumerate}

Proof of the paths (1+2) and (1--2) are straightforward.

Let us proof the path (2..3).
For it let us assume that the condition assigned to the control point 2 is true.
It implies that before exercise of $mixSQR$ the precondition in (\ref{mufixSQRass}) is true;
due to the correctness of the assertion (\ref{mufixSQRass}), the algorithm $mixSQR$ terminates
and upon its termination the postcondition in (\ref{mufixSQRass}) is true also.
Remark also that conjuncts \ref{Y} and \ref{Z} remains true since $mixSQR$ doesn't change neither $Y$ nor $A$.
Hence the condition assigned to the control point 3 is true.
It finishes the proof of the path (2..3).

For proving the last path (3..4)
firstly let us prove  that the condition in the control point (3) implies the next two properties:
\begin{equation}\label{3impr1}
\frac{Z}{2}=\lfloor\frac{Exp(A)}{2}\rfloor,
\end{equation}
\begin{equation}\label{3impr2}
\begin{array}{c}
|X - \sqrt{if\ odd(Exp(A))\ then\ Man(A)\times\beta_F\ else\ Man(A)}|\ < \\
\hspace*{8cm} <\ \big(Eps+\frac{\delta_T}{2\sqrt{\beta_F}}\big).
\end{array}
\end{equation}
The first  property (\ref{3impr1}) directly follows from the condition (\ref{Z}).
The prove of the second one follows below:
\begin{description}
  \item[$Exp(a)$ is even:] The radicand and the variable $Y$ (in algorithm $mixSQR$)
  both equal $Man(A)$; hence
  \begin{center}
   $|X - \sqrt{if\ odd(Exp(A))\ then\ Man(A)\otimes\beta_F\ else\ Man(A)}|\ =$ \hspace*{\fill} \\
   $=\ |X - \sqrt{Y}|\ <$ (because assertion (\ref{mufixSQRass}) is correct)$<$\\
   \hspace*{\fill}  $<\ Eps\ <\  \big(Eps+\frac{\delta_T}{2\sqrt{\beta_F}}\big)$.
  \end{center}
  \item[$Exp(a)$ is odd:] The radicand equals $Man(A)\times\beta_F$ and the variable $Y$ equals $Man(A)\otimes\beta_F$;
  hence
  \begin{center}
   $|X - \sqrt{if\ odd(Exp(A))\ then\ Man(A)\times\beta_F\ else\ Man(A)}|\ =$ \hspace*{\fill} \\
   $=\ |X - \sqrt{Man(A)\times\beta_F}|\ <$ \hspace*{\fill} \\
    $<\ |X - \sqrt{Man(A)\otimes\beta_F}|\ +\ |\sqrt{Man(A)\otimes\beta_F} - \sqrt{Man(A)\times\beta_F}|\ <$\\
   (due to correctness of the  assertion (\ref{mufixSQRass}))\\
   \hspace*{\fill} $<\ Eps\ + |\sqrt{Man(A)\otimes\beta_F} - \sqrt{Man(A)\times\beta_F}|<$\\
(because of properties of the fix-point arithmetic, where $|\Delta|<\frac{\delta_T}{2}$)\\
\hspace*{\fill} $<\ Eps\ + |\sqrt{Man(A)\times\beta_F + \Delta} - \sqrt{Man(A)\times\beta_F}|<$\\
   (because of Taylor expansion of $\sqrt{t+\Delta}$ as a series $\sqrt{t} + \frac{\Delta}{\sqrt{t}} + \dots$)
\hspace*{\fill} $<\ Eps\ + |\sqrt{Man(A)\times\beta_F }+ \frac{\Delta}{\sqrt{Man(A)\times\beta_F}} + ... - \sqrt{Man(A)\times\beta_F}|<$\\
\hspace*{\fill} $<\ Eps\ + |\frac{\Delta}{\sqrt{Man(A)\times\beta_F}} + ...|<$\\
(because the Taylor expansion of  $\sqrt{t+\Delta}$ is  an alternating series) \\
\hspace*{\fill} $<\ Eps\ + |\frac{\delta_T}{2\sqrt{Man(A)\times\beta_F}}|<$\\
\hspace*{\fill} $<$(because $1<Man(A)$)$<\ Eps\ + \frac{\delta_T}{2\sqrt{\beta_F}}$.  \end{center}
\end{description}
As soon as it is proved that the properties (\ref{3impr1}) and (\ref{3impr2}) are valid in the control point (3),
the proof of the path (3..4) become trivial.

It finishes pen-and-paper verification of specified algorithm computing approximations for the square root with floating-point arithmetic.

\section{Conclusion}\label{Concl}
Let us summarise the content of the paper.

Firstly we take a very standard Newton method to compute square root, present is as an iterative algorithm $SQR$,
specify it by Hoare total correctness assertion, and prove its validity in the case when input argument is greater than 1,
accuracy is positive, and ``computer'' is precise (i.e. all computations are done in mathematical real numbers);
the upper bound of loop iterations of the algorithm $SQR$ is logarithmic.

Next we improve the algorithm $SQR$ by using an auxiliary function to compute better initial approximations for square roots
(it results in the algorithm $ISQR$) and then suggest a for-loop-based algorithm $FSQR$ that uses the same auxiliary function,
computes a lower bound for the number of iterations that is sufficient to achieve the specified accuracy;
both algorithms $ISQR$ and $FSQR$ work with precise arithmetic, but we prove that $FSQR$ achieves better accuracy than $ISQR$,
and can achieves better accuracy if to increase the number of the loop iterations.

Then we convert for-loop-based algorithm with precise arithmetic $FSQR$ into algorithm $fixSQR$ with fix-point arithmetic,
specify it by total correctness assertion and prove its validity by adjustment of its runs with runs of $FSQR$ with the same input data.
Another specifics of the algorithm $fixSQR$ is use a look-up table (arrange as an array)
for upper approximations of square roots and rounding-up function.

Use of a machine fix-point arithmetic instead of the precise arithmetic results in situation that more iterations of the loop
doesn't always improve accuracy in contrast to $FSQR$. Due to this reason  we suggest an other algorithm $mixSQR$ that
is a specialised version of the algorithm $fixSQR$.

Finally we use the algorithm $mixSQR$ as a subroutine in algorithm $fltSQR$ that computes approximations of the square root
function in floating-point arithmetic. For this we assume that each floating-point number is represented as its mantissa and exponent,
and both --- the mantissa and exponent --- are fix-point numbers. We specify the algorithm by a total correctness assertion and prove its
correctness (basing on the correctness of the algorithm $mixSQR$).

All proofs in this paper are human-driven and oriented pen-and-paper proofs.
So the next topic of our project is to validate all these proofs with aid of some automated proof-assistant.
We are going to use ACL2 due to  industrial strength of this proof-assistant \cite{Grohoski17}
for platform-specific verification of the standard mathematical functions
(but don't rule out alternatives to this assistant).

Nevertheless remark that we attempt and present in this paper an approach that we call platform-independent.
Also remark that we don't attempt to build an axiomatization of an ``abstract'' machine
(fix-point or floating-point) arithmetics.
Instead we just make several explicit assumptions about machine arithmetic (and how it relates to the precise arithmetic)
that are sufficient to validate specifications and algorithms with machine arithmetic by using
its relations with specifications and algorithms with precise arithmetics.
We believe that our assumptions about machine arithmetic are valid for many platforms and
they are easy to check.
Remark that if a platform's machine arithmetic meets these assumptions then
properties of the algorithms $mixSQR$ and $fltSQR$ exercised on this platform
are specified by total correctness assertions (\ref{mufixSQRass}) and (\ref{fltSQRass}) respectively.

Let us group together and list in one place our assumptions about fix-point and floating-point machine arithmetic
that we introduce in the subsection \ref{FixPointArith} and the section \ref{fltSect}) and use in this paper:
\begin{description}
  \item[Fix-point arithmetic:] We understand fix-point numeric data type $T$ as follows:
\begin{itemize}
\item the set of values $Val_T$ is a finite subset of mathematical reals $\mathbb{R}$
such that
\begin{itemize}
\item it comprises all reals in some finite range $[-\inf_T,\sup_T]$, where $\inf_T>2$, $\sup_T>2$, with some fixed step $\frac{1}{2}>\delta_T>0$,
\item and includes all integer numbers $Int_T$ in this range $[-\inf_T,\sup_T]$;
\end{itemize}
\item legal binary arithmetic operations are
\begin{itemize}
\item addition and subtraction;
if not the range overflow exception then these operations are precise: they equal to the standard mathematical operations
assuming their mathematical  results fall in the range $[-\inf_T,\sup_T]$
(and due to this reason are denoted as $+$ and $-$);
\item multiplication $\otimes$ and division $\oslash$;
these operations are approximate but correctly rounded in the following sense:
for all $x,y\in Val_T$
\begin{itemize}
\item if $x\times y\in Val_T$ then $x\otimes y = x\times y$;
\item if $x/y\in Val_T$ then $x\oslash y = x/y$;
\item if $x\times y\in [-\inf_T,\sup_T]$ then $|x\otimes y - x\times y|<\delta_T/2$;
\item if $x/y\in [-\inf_T,\sup_T]$ then $|x\oslash y - x/y|<\delta_T/2$.
\end{itemize}
\end{itemize}
\item legal binary relations are equality and all standard inequalities;
these relations are precise, i.e. they equal to the standard mathematical relations
(and due to this reason are denoted as $=$, $\neq$, $\leq$, $\geq$, $<$, $>$).
\end{itemize}
  \item[Floating-point arithmetic] We understand floating-point numeric data type $F$ as follows:
\begin{itemize}
\item the set of values $Val_F$ is a finite subset of mathematical reals $\mathbb{R}$
that comprises some reals in some finite range $[-\inf_F,\sup_F]$, where $\inf_F,\sup_F>2$ and $\{0,\inf_F,\sup_F\}\subseteq Val_F$;
\item there are two unary operations $Man:Val_F\rightarrow Val_T$ (called \emph{mantissa}),
$Exp:Val_F\rightarrow Int_T$ (called \emph{exponent}), and an integer constant $\beta_T\in Int_F$
(called \emph{exponent base} or just \emph{base})
such that for all positive $x\in Val_F$
\begin{itemize}
\item $1< Man(x)< \frac{sup_T}{\beta_F}$;
\item if $inf_T$ is odd then $-inf_T< Exp(x)$ else $-inf_T\leq Exp(x)$;
\item $x= Man(x)\times \beta_F^{Exp(x)}$.
\end{itemize}
\end{itemize}
\end{description}

Let us remark that our fix-point and floating-point numeric types
are \emph{internal} or \emph{instant} types in the following sense:
\begin{itemize}
\item a program language provides numeric user-types \texttt{integer}, \texttt{real}, etc.
(they may be \texttt{int} and/or \texttt{long int}, \texttt{float} and/or \texttt{double} in C,
or \texttt{integer} and \texttt{real} in Pascal, etc.) with type-, implementation-, and platform-dependent
the  \emph{unit of least precision} (or \emph{unit in the last place}) $ulp_\tau\in \mathbb{R}$,
where $\tau$ is a ``complex parameter'' (type, implementation, platform);
\item our fix-point type $T$ and floating-point type $F$ are types for \emph{microprograms} to implement
algorithms $mixSQR$ and $fltSQR$ in such a way that there exist values $\varepsilon>0$ and $s>0$ for
variables $Eps$ and $Stp$ that guaranty \emph{exact} accuracy of the implemented $mixSQR$ and $fltSQR$,
(i.e. $\frac{ulp_\tau}{2}>\varepsilon$ in the case of \texttt{integer} and algorithm $mixSQR$, and
$\frac{ulp_\tau}{2}>\varepsilon +\frac{\delta_T}{2\sqrt{\beta_F}}$ in the case of \texttt{real} and algorithm $fltSQR$).
\end{itemize}

Finally let us mention one more research topic ---
to find an ``optimal balance'' between size
of the array $root$ with initial upper approximations for square roots for selected arguments,
number of iterations of the loop in the algorithm $finSQR$,
and accuracy of the square root approximation:
if $\varepsilon$ and $s$ are values of the variables $Eps$ and $Stp$ then the array size is $\frac{\sup_T}{s}$,
number of iterations may be any $n\geq \big(1+\lceil\log_2\frac{s}{\varepsilon}\rceil\big)$,
and accuracy $|X - \sqrt{Y}|$ is less than $\big(\frac{s}{2^{n}}\ +\ n\times\delta_T\big)$.

\newpage

\appendix

\section{Specification of the \texttt{sqrt} function}\label{sqrt-spec}
The C reference portal  at \url{en.cppreference.com/w/c}
specifics the the square root function \texttt{sqrt} \cite{SqrtCref} as it represented below.

\vspace*{2ex}
{\small
\begin{tabular}{|l|}
\hline
\hspace*{11cm} \\
\noindent
C  Numerics  Common mathematical functions \\
\textbf{\normalsize sqrt, sqrtf, sqrtl} \\
Defined in header $<$math.h$>$\\
float       sqrtf( float arg ); (1)	(since C99)\\
double      sqrt( double arg ); (2)	\\
long double sqrtl( long double arg ); (3)	(since C99) \\
Defined in header $<$tgmath.h$>$ \\
$\sharp$define sqrt( arg ) (4)	(since C99)\\
1-3) Computes square root of arg.\\
4) Type-generic macro: If arg has type long double, sqrtl is called. \\
Otherwise, if arg has integer type or the type double, sqrt is called. \\
Otherwise, sqrtf is called. If arg is complex or imaginary, \\
then the macro invokes the corresponding complex function \\
(csqrtf, csqrt, csqrtl).\\
\textbf{Parameters}\\
arg - floating point value\\
\textbf{Return value}\\
If no errors occur, square root of arg ($\sqrt{arg}$), is returned.\\
If a domain error occurs, an implementation-defined value is returned \\
(NaN where supported). \\
If a range error occurs due to underflow, the correct result (after rounding)\\
is returned.\\
\textbf{Error handling}\\
Errors are reported as specified in math\_errhandling.\\
Domain error occurs if arg is less than zero.\\
If the implementation supports IEEE floating-point arithmetic (IEC 60559),\\
$\bullet$ If the argument is less than $-0$, FE\_INVALID is raised and NaN is returned.\\
$\bullet$ If the argument is $+\infty$ or $\pm$0, it is returned, unmodified.\\
$\bullet$ If the argument is NaN, NaN is returned \\
\textbf{Notes} \\
sqrt is required by the IEEE standard be exact. \\
The only other operations required to be exact are the arithmetic operators \\
and the function fma. After rounding to the return type \\
(using default rounding mode),\\
the result of sqrt is indistinguishable from the infinitely precise result. \\
In other words, the error is less than $0.5$ ulp. \\
Other functions, including pow, are not so constrained.\\
\hspace*{11cm} \\
\hline
\end{tabular}
}


\begin{thebibliography}{9}

\bibitem{AyadMarche10}
Ayad A., March\'{e} C.
Multi-prover verification of floating-point programs //
Lecture Notes in Artificial Intelligence. 2010. Vol. 6173. P.127--141.

\bibitem{Barret89}
Barret G. 1989. Formal Methods Applied to a Floating-Point Number System // IEEE Trans. Softw. Eng. 1989. Vol. 15(5). P.611--621.

\bibitem{BrainTinelliRuemmerWahl15}
Brain M., Tinelli C., Ruemmer Ph., and Wahl T.
An Automatable Formal Semantics for IEEE-754 Floating-Point Arithmetic //
Proceedings of the 2015 IEEE 22nd Symposium on Computer Arithmetic (ARITH '15). IEEE Computer Society. 2015. P160--167.

\bibitem{FergusonBinghamErkokHarrisonLeslie-Hurd17}
Ferguson W.E. (Jr), Bingham J., Erkok L., Harrison J.R., Leslie-Hurd J.
Digit serial methods with applications to division and square root (with mechanically checked correctness proofs) //
2017. Eprint arXiv:1708.00140. \url{https://arxiv.org/abs/1708.00140}. (Visited December 19, 2017.)

\bibitem{Gamboa97}
Gamboa R.A. Square Roots in Acl2: a Study in Sonata Form // Technical Report. University of Texas at Austin, USA. 1997.

\bibitem{Gries81}
Gries D. The Science of Programming. Springer-Verlag, 1981.

\bibitem{Harrison99}
Harrison J. A Machine-Checked Theory of Floating Point Arithmetic //
Lecture Notes in Computer Science. 1999. Vol. 1690. P. 113--130.

\bibitem{Harrison00a}
Harrison J. Formal Verification of Floating Point Trigonometric Functions //
Lecture Notes in Computer Science. 2000. Vol. 1954. P.217-233.

\bibitem{Harrison00b}
Harrison J. Formal Verification of IA-64 Division Algorithms //
Lecture Notes in Computer Science. 2000. Vol. 1869. P. 233--251.

\bibitem{Harrison00c}
Harrison J. Floating Point Verification in HOL Light: The Exponential Function //
Formal Methods System Design. 2000.  Vol. 16(3). P. 271--305.

\bibitem{Harrison03}
Harrison J. Formal Verification of Square Root Algorithms //
Formal Methods in System Design. 2003. Vol. 22(2). P.143--153.

\bibitem{Hoare03}
Hoare C.A.R. The Verifying Compiler:
A Grand Challenge for Computing Research //
Lecture Notes in Computer Science. 2003. Vol. 2890. P. 1--12.

\bibitem{Grohoski17}
Grohoski G. Verifying Oracle's SPARC Processors with ACL2. Slides of the Invited talk for 14th International Workshop on the
ACL2 Theorem Prover and Its Applications.
\url{http://www.cs.utexas.edu/users/moore/acl2/workshop-2017/slides-accepted/grohoski-ACL2_talk.pdf}.
(Visited December 19, 2017.)

\bibitem{Gutowski}
Gutowski M.W. Power and beauty of interval methods. 	
arXiv:physics/0302034 [physics.data-an]. \url{http://arxiv.org/pdf/physics/0302034.pdf}.
(Visited December 19, 2017.)

\bibitem{Kochan05b}
Kochan S.G. Programming in C: A Complete Introduction to the C Programming Language. Functions Calling Functions at p.131.
Sam's Publishing, 2005 (3rd  Edition).

\bibitem{Kuliamin07}
Kuliamin V. Standardization and Testing of Mathematical Functions //
Programming and Computer Software. 2007. Vol. 33, n. 3. P. 154--173.

\bibitem{Kuliamin10}
Kuliamin V.V.
Standardization and Testing of Mathematical Functions in floating point numbers //
Lecture Notes in Computer Science. 2010. Vol. 5947. P. 257--268.

\bibitem{Monniaux08}
Monniaux D. The pitfalls of verifying floating-point computations //
ACM Transactions on Programming Languages and Systems. 2008. Vol. 30, n. 3. P.1--41.

\bibitem{Muller05}
Muller J.-M. Elementary Functions: Algorithms and Implementation. Birkhauser, 2005.

\bibitem{SawadaGamboa02}
Sawada J., Gamboa R. Mechanical Verification of a Square Root Algorithm Using Taylor's Theorem //
Lecture Notes in Computer Science. 2002. Vol. 2517. P. 274--291.

\bibitem{Shelehov10}
Shelihov V.I. Verification and synthesis of efficient programs for standard functions \texttt{flor}, \texttt{isqrt} and \texttt{ilog2}
using predicate programming technology //
Proceedings of 12 Int. Conf. on Control and modelling of complex systems. Samara: Samara Science Center of Russian Academy of Science.
2010. P.622-630. (In Russian.)

\bibitem{Shilov15}
Shilov N.V. On the need to specify and verify standard functions //
The Bulletin of the Novosibirsk Computing Center
(Series: Computer Science). 2015. n.38, p.105--119.

\bibitem{ShilovPromsky16}
Shilov N.V., Promsky A.V. On specification andd verification of standard mathematical functions //
Humanities and Science University Journal. 2016. n.19, p.57--68.

\bibitem{SiddiqueHasan14}
Siddique U., Hasan O. On the Formalization of Gamma Function in HOL //
2014. J. Autom. Reason. Vol. 53(4). P. 407--429.

\bibitem{SqrtCref}
C refernce. Sqrt, sqrtf, sqrtl.
\url{http://en.cppreference.com/w/c/numeric/math/sqrt}. (Visited December 19, 2017.)

\bibitem{IEEEstandard}
IEEE 754-2008.
\url{http://ieeexplore.ieee.org/document/4610935}. (Visited December 19, 2017.)

\bibitem{IECstandard}
ISO/IEC/IEEE 60559:2011.
Information technology \texttt{--} Microprocessor Systems
\texttt{--} Floating-Point arithmetic
\url{http://www.iso.org/iso/iso_catalogue/catalogue_tc/catalogue_detail.htm?csnumber=57469}.
(Visited December 19, 2017. In Russian.)


\bibitem{Roscosmos}
 Roskosmos called the reason of unsuccessful start from the East spaceport
\url{https://news.mail.ru/politics/31931345/}. (Visited De\-cem\-ber 19, 2017.)

\end{thebibliography}
\end{document}